# EPITAXIAL GROWTH AND STABILIZATION OF PEROVSKITE PHASE EuNiO$_3$ THIN FILMS THROUGH RF SPUTTERING

[a.] Centre For Nano Science & Engineering, Indian Institute of Science, Bengaluru, India 560012

Prashanth S [a], Binoy Krishna De [a], Shubham Kumar Parate [a], Kartick Biswas [a] and Pavan Nukala [a*]

*Corresponding author email: pnukala@iisc.ac.in

Phase change materials (PCMs) that show volatile resistive switching can be implemented to emulate neuronal oscillators. Charge transfer insulators, ReNiO$_3$ with Re = Rare earth metal (Pr, Nd, Sm, Eu...) present a family of PCMs that show tunable metal-insulator transition (MIT) in a wide range of temperatures. Interestingly, the MIT can be tuned via chemical doping and/or strain engineering. EuNiO$_3$ in particular can be a desirable choice for oscillator devices given that bulk transition temperature $T_{MI}$ ~190$^0$ C, which is well above the room temperature, avoiding issues with crosstalk, yet not too high to still promise low power application. We report a strategy to stabilize high-quality epitaxial EuNiO$_3$ thin films using a scalable reactive RF sputtering process and post-annealing treatment, by systematic variation of oxygen partial pressure and annealing temperatures on two different substrates. We show that growth at low or no oxygen partial pressures yields amorphous samples. On the other hand, at higher pressures, crystallinity is improved but Ruddlesden's- Popper's (RP) phases [$A_{n+1}B_nO_{3n+1}$] and corresponding faults stabilize. Post-annealing improves the crystallinity in all the samples and triggers the transformation of the RP phases/faults towards the perovskite phase. We performed a combination of operando XRD and transport measurements on our films and determined the transition temperature of perovskite phase samples grown on LAO and LSAT to be ~300$^0$ C and ~250$^0$ C, respectively. We believe that in-plane compressive strain (LAO:-0.76%) will decrease the Ni-O-Ni bond angles in-plane, which results in an increase in $T_{MI}$ compared to bulk. On the other hand, in-plane tensile strain (LSAT:2.1%) also decreases Ni-O-Ni bond angles out-of-plane resulting in an increase of $T_{MI}$ compared to bulk, which is however offset by the creation of oxygen vacancies owing to their lowered formation energies. Noticeably, the films stabilized in RP phases did not show any transition.

## Introduction

Strongly electron-correlated materials are a class of functional materials with complex interplay of spin, lattice, orbital, and charge degrees of freedom. [1,2] A class of correlated materials, Mott insulators show a metal-to-insulator transition (MIT), which can be exploited to construct neuromorphic oscillators (neuristors) that operate at room temperature. VO$_2$ ($T_{MI}$ ~ 67$^0$ C) and NbO$_2$ ($T_{MI}$ ~ 808$^0$ C) [1] are the preferential choice for such applications. However, neuristors based on VO$_2$ can easily suffer from thermal crosstalk, whereas those based on NbO$_2$ require very high current densities (and hence power) to generate oscillations. It is thus ideal to have materials with $T_{MI}$ ~150$^0$ C to 400$^0$ C for these applications. Rare earth nickelates (ReNiO$_3$) are charge transfer insulators that offer a wide range of $T_{MI}$ depending on the choice of Re. [1,2] Additionally, the $T_{MI}$ can be tuned with chemical doping, strain, pressure, and stoichiometry. [3–7] Though, epitaxial films with Re = Pr, Nd, and Sm have been thoroughly explored in various works, [8–16] there are very few reports on the ReNiO$_3$ thin films with Re=Eu, Gd which have a lower ionic size and larger $T_{MI}$ [17,18]. In particular, EuNiO$_3$ (ENO) has a $T_{MI}$ of ~190$^0$ C in bulk, which is in the desired temperature range. However, fundamental aspects such as the effects of strain and oxygen vacancies on $T_{MI}$ in this system have not been investigated.

In this work, we used reactive RF sputtering and systematically varied oxygen partial pressure and post-annealing temperatures to study the evolution of both epitaxial perovskite and Ruddlesden's Popper's (RP) phases of Europium Nickelate. Through a combination of operando XRD and transport measurements, we determine the transition temperatures of various films while understanding the role of defects in addition to strain in tuning the $T_{MI}$.

# Experimental

## Methods and Synthesis
### Substrate Treatment methods:

LaAlO$_3$ (LAO) and [LaAlO$_3$]$_{0.3}$[Sr$_2$TaAlO$_6$]$_{0.7}$ (LSAT) single crystalline substrates with lattice constants of 3.76 Å and 3.89 Å, were used respectively for ENO film deposition. The substrates were sonicated in acetone, and IPA and rinsed in DI water to remove any particle contaminants. After solvent treatment, the LAO substrates were post-annealed in a zone furnace in O$_2$ atmospheres at 1200$^0$ C for 5 hours to create an atomically flat surface exposing the terraces as shown by AFM (Supplementary Figure S1).

### Sputtering and post-annealing methods:

A 3-inch diameter sputtering target which has a mixed composition of Eu$_2$O$_3$, NiO, and EuNiO$_3$, (target from *STANFORD ADVANCED MATERIALS*) was used. The chamber is pumped to low pressures of 2 x 10$^{-6}$ mbar. Before every deposition, the target was pre-sputtered for 30 minutes with pure Argon to remove contaminants and another 30 minutes with O$_2$ to condition the target. During the entire pre-sputter process the shutter was closed. The target-to-substrate distance of 60 mm, RF power of 100 W, and substrate temperature of 700$^0$ C were kept constant for the deposition conditions. Argon to oxygen flow rates were systematically varied to optimize the ideal growth conditions for epitaxial films. Here we predominantly discuss results on three sets of representative samples grown on LSAT: S1 to S3 and three sets on LAO (L1 to L3), each grown at different oxygen partial pressures. Table 1 shows the sputtering parameters and conditions for growth of samples S1 to S3 (L1 to L3). Post sputtering, the samples were annealed at 800$^o$ C in a zone furnace with varying durations of 12 hours to 36 hours respectively. During annealing, a constant O$_2$ flow of 2 lpm (liters per minute) was maintained.

**Table 1: Sputtering Parameters of EuNiO3 grown on LAO(L1-L3) and LSAT(S1-S3)**

| SAMPLE | ARGON FLOW (SCCM) | O2 FLOW (SCCM) | O2 PARTIAL PRESSURE (MBAR) | DEP. RATE (NM/MIN) |
|---|---|---|---|---|
| S1 (L1) | 50 | - | - | 6.0 |
| S2 (L2) | 50 | 15 | 0.0010 | 2.0 |
| S3 (L3) | 20 | 40 | 0.0024 | 0.5 |

### Characterization Methods:

X-ray Diffraction studies were performed using a Cu source of wavelength 1.54 Å. The structural imaging was done using HAADF-STEM on Thermofisher TITAN Themis 300 at 300kV with 24 mrad convergence angle at 160 mm camera length. Compositional analysis was carried out with Titan Themis 300 equipped with SuperGX quad detectors. The temperature-dependent resistivity of the thin films was measured in the linear point probe configuration, using Keithley 2425 source meter and Keithley 2002 multimeter. The sample temperature was controlled using a Linkam TS1000 temperature controller. Temperature-dependent resistivity measurements were performed in the temperature sweep mode with a 3K/min temperature sweep rate with an applied current of 1 µA.

## Results and discussion

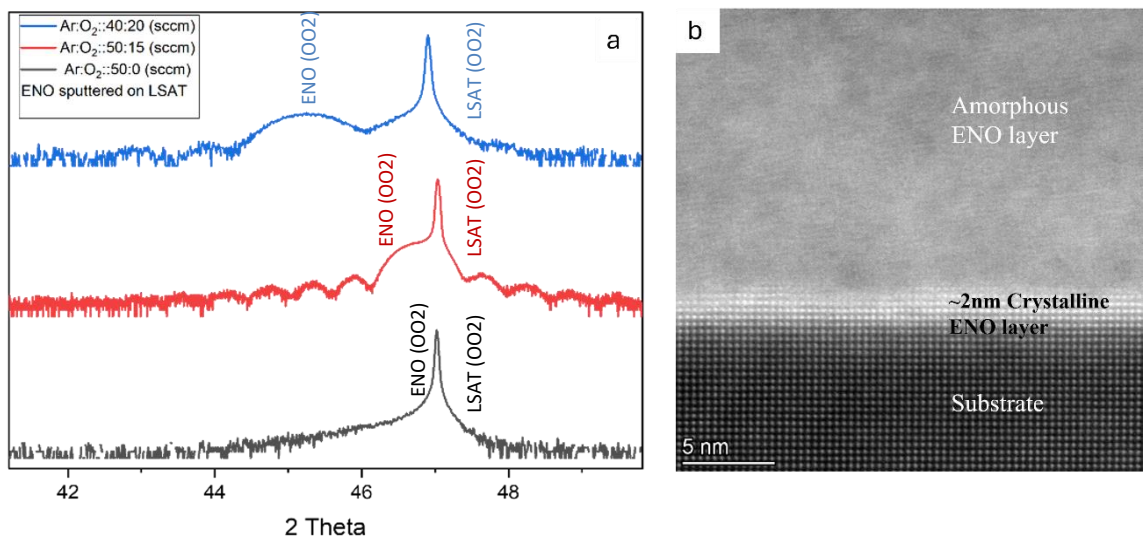

**Fig 1a**) XRD measurements of ENO on LSAT sputtered under different Ar and O2 flow rates. **Fig1b**) HAADF-STEM image of ENO sputtered with only Argon.

Theta-2 theta XRD scans obtained from samples S1 to S3 are displayed (fig 1a). S1 and L1 (shown in supplementary figure S2), grown under pure Ar atmosphere do not show any Bragg peaks, and corresponding HAADF-STEM images (Fig. 1b) reveal that mostly the film is amorphous, except for a coherent 3-4 layers (~2nm layer with bright contrast) of epitaxial film at the interface with the substrate. With increasing $O_2$ partial pressure, we observe the emergence of a Bragg peak, which shifts to lower 2 theta values of 46.70(3)$^0$ and 45.2(5)$^0$ (or larger c-parameter 3.894(6) Å and 4.02(8) Å), respectively on samples S2 and S3. The Laue oscillations are indicative of the good crystalline quality of these films. L1 to L3 grown on LAO (see Supplementary Figure S2) also shows similar trends. Following the work of Pan et.al[14], we posit that the increase in lattice parameter is a result of the formation of non-stochiometric (oxygen deficient) $EuNiO_3$ thin films. Our HAADF-STEM imaging results (presented later) indeed confirm that the larger lattice parameters emanate from the formation of RP phases and defects. Thus, although higher $O_2$ partial pressure yields better crystallinity, it favours the formation of RP phases in competition to the perovskite phase.

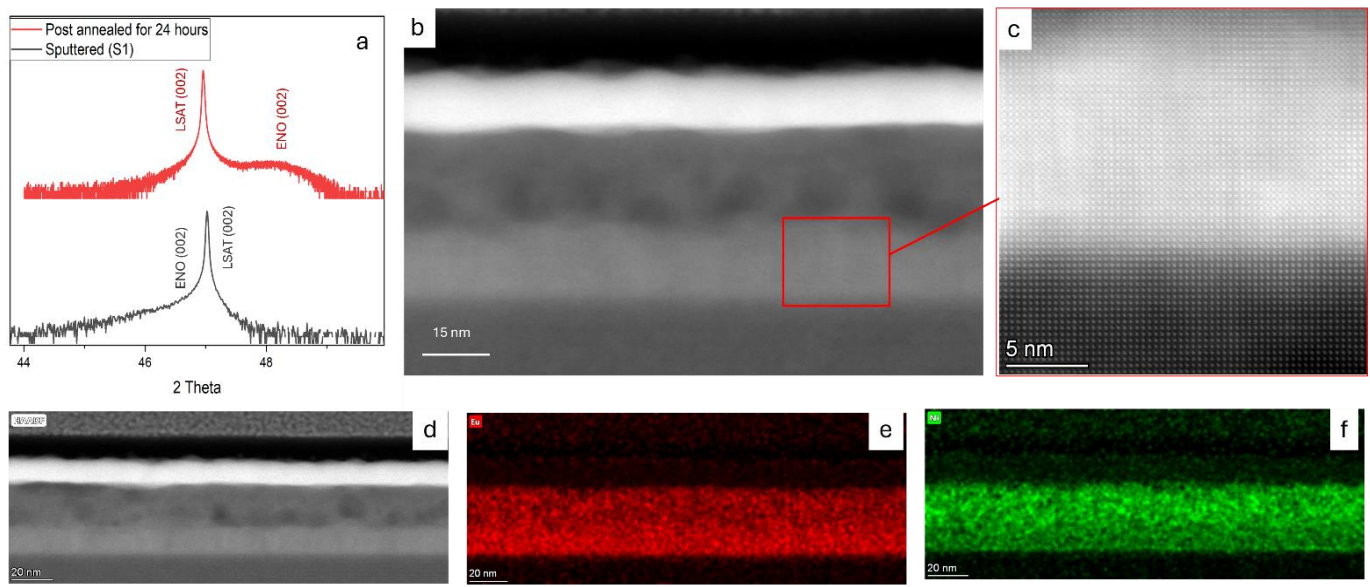

**Fig 2a**) XRD measurements of S1 post sputtered and post annealed after 24 hours. **Fig 2b**) HAADF-STEM image of L1 post annealed for 24 hours **Fig 2c**) Crystallization of L1 post annealing. **Fig 2d**) HAADF image of the showing crystalline and amorphous region. **Fig 2e**) Corresponding EDS spectra from the HAADF image (Fig 2d) of Eu in the amorphous and crystalline region and **fig 2f**) similarly EDS spectra of Ni.

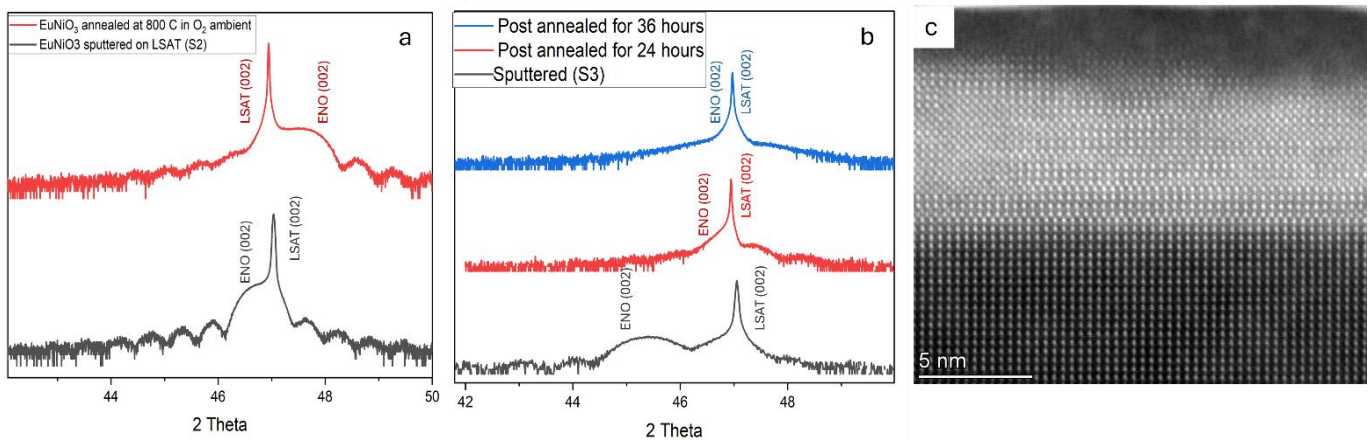

**Fig 3a)** XRD data of ENO on LSAT (S2) post-sputtered and post-annealed. The peak shift towards the right at 2theta ~ 47.7400 shows that the film is stabilized to the perovskite phase. Post-annealing shows that ENO is stabilized in the perovskite phase. **Fig 3b)** XRD data of ENO on LSAT (S3) post-sputtered and post-annealed shows that prolonged annealing durations stabilize the film to RP phases. **Fig 3c)** HAADF-STEM image showing RP phases and faults of ENO on LAO (L3)

We note that growth rates reduce with an increase in $O_2$ partial pressure (Table 1). Owing to the lower mass of oxygen (and oxygen ions), the sputter yield is less at higher oxygen pressures, suggesting that the plasma is less energetic, compared to the plasma arising from pure Ar sputtering. This possibly results in the formation of thermodynamically stabilized structures (in this case, crystalline RP phases), whereas lack of oxygen increases the sputter yield and plasma energy, kinetically stabilizing amorphous $EuNiO_3$. We posit that the thin layer of perovskite $EuNiO_3$ at the interface in S1 is a result of interface energy minimization, which prefers coherent interfaces (fig 1b).

Next, in order to crystallize and stabilize the perovskite structure in all the samples, we post-annealed them for varying times and followed their crystallization process through both XRD and HAADF-STEM measurements (fig 2a & 2b). For S1, it is observed after 24 hours of annealing in a continuous stream of $O_2$ flow there is an improvement in the crystallization with the bottom 15 nm close to the substrate interface crystallizing into a perovskite phase. We observe the same trend in films grown on LAO (supplementary Figure S3). The elemental maps acquired using energy dispersive spectroscopy (fig 2d-f) of the entire annealed stack, clearly show that the amorphous ENO is richer in Ni (supplementary fig S4).

When S2 is post-annealed using similar conditions, we see that the Bragg peak at 2-theta = $46.70(3)^0$ (peak A) shifts right to 2-theta=$47.468(5)^0$ (fig 3a), indicating a reduction in the c-lattice parameter from 3.894(6) Å to 3.806(7)Å. This also corresponds to a gradual formation of the perovskite phase from the RP phases across the entire film. Post annealing of S3 (fig 3b) shows similar trends, but an incomplete conversion of RP to the perovskite phase (supplementary figure S5). To check whether increasing the annealing time will complete the transition, we further annealed S3 for 24 and 36 hours, respectively. Although there is a substantial shift of peak A towards the right, complete transition is not achieved. HAADF-STEM data (fig 3c) shows RP phases in these films, suggesting a transition of RP phases towards higher "n" compounds in the homologous series, $A_{n+1}B_nO_{3n+1}$ with n>2. In S1 and S2, we believe that supplying excess oxygen in addition to the presence of some excess Ni leads to the complete transformation of RP phases to the perovskite phase. However, S3 crystallizes in an RP phase compound, lower in the homologous series (lower n in $A_{n+1}B_nO_{3n+1}$). Hence, we believe that annealing not just in an oxygen atmosphere, but also excess Ni should help a faster conversion into perovskite phase.

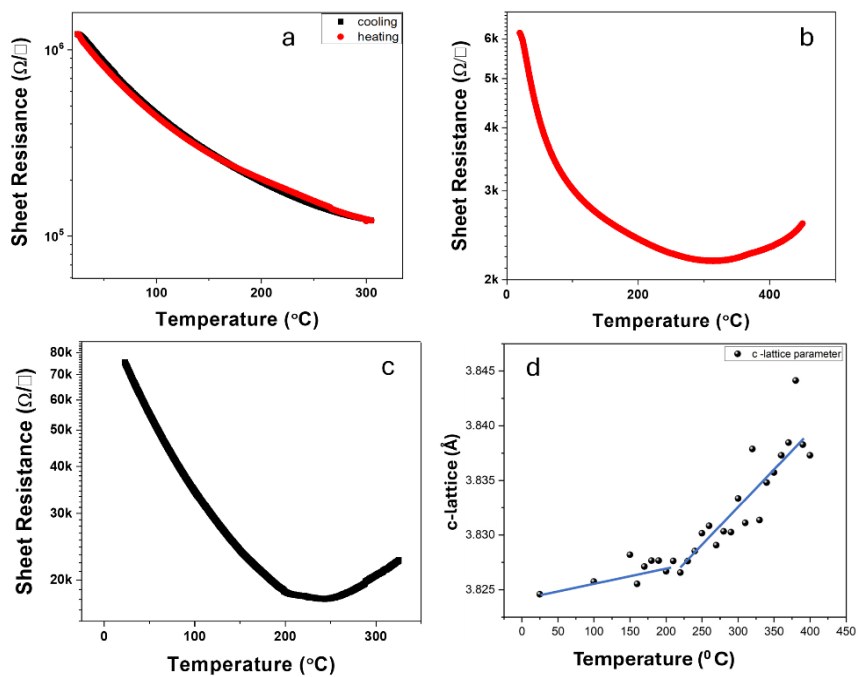

**Fig 4a**) Transport measurements of RP phase ENO on LSAT showing insulating behavior. **Fig 4b**) Transport measurements of ENO on LAO. **Fig 4c**) Transport measurements of ENO on LSAT. **Fig 4d**) In situ temperature dependent XRD of ENO on LSAT.

Next, we present transport measurements to compare possible electronic transitions occurring across different samples. Samples with predominantly RP phase (annealed S3, L3) show an insulating behaviour with the resistance decreasing with an increase in temperature from room temperature to $300^0$ C (Fig 4a). However, samples with predominantly perovskite phase show insulator-metal transition. ENO on LAO shows an electronic transition at ~$300^0$ C (fig 4b) and on LSAT shows a transition at ~$250^0$ C (fig 4c). We also confirm that this transition is indeed associated with structural transitions through operando temperature-dependent XRD (fig 4d), where we followed the evolution of the c lattice parameter using (002) Bragg peak at 2 theta = $47.74^0$ as a function of temperature. The discontinuous slope change in temperature evolution of the c lattice parameter at ~250°C indicates the structural modification associated with the MIT observed in the transport study (Fig 4a).

The lattice parameters of ENO films measured from HAADF-STEM analysis on LAO are a=3.748 Å (±0.0213) and c=3.798 Å (±0.0205), which are in-plane compressively strained by – 0.76 % compared to the bulk values in orthorhombic phase. It is known nickelates under compressive strain show an increase in the $T_{MI}$ owing to a reduction in Ni-O-Ni bond angle away from 180° [16]. However, LSAT substrates impose an in-plane tensile strain on ENO (2.1%) which also results in a reduction of Ni-O-Ni bond angle in the out-of-plane direction, which enhances the transition temperature from bulk. However, in-plane compressive strain is also conducive to the creation of oxygen vacancies, which is reflected as a larger volume of the unit cell with a=3.897 Å (±0.03904), c=3.813 Å(±0.0228). Oxygen vacancies in epitaxial $NdNiO_3$ films are known to reduce the transition temperature[16], which is also consistent with lesser TMI for ENO on LSAT compared to that on LAO.

## Conclusion

In this work, we demonstrate the optimal synthesis of epitaxial $EuNiO_3$ in the perovskite phase by systematic variation of oxygen partial pressure during RF sputtering and post-annealing conditions on LAO and LSAT substrates. Higher oxygen pressures stabilize RP phases, while smaller pressures stabilize amorphous perovskites. Post annealing helps both sets of samples driving them towards an epitaxial perovskite phase. We show that compressively strained ENO on LAO shows an enhanced $T_{MI}$ of $300^0$ C compared to bulk, which slightly depresses on LSAT to ~$250^0$ C owing to the presence of oxygen vacancies stabilized by tensile strain. Our work demonstrates a systematic procedure to synthesize desired robust ReNiO$_3$ epitaxial films (charge transfer insulators) using a scalable RF sputtering process.

## Author contributions

Ideation began with discussions between Pavan Nukala (PN) and Prashanth S (PS). The design of experiments was carried out by PN and PS. The RF sputtering, annealing, XRD and in situ XRD measurements were carried out by PS. Binoy Krishna De (BK) carried out transport measurements of the samples. BK helped in the data analysis of the XRD in situ measurements. Shubham Kumar Parate (SK) and Kartick Biswas (KB) prepared the TEM sample. SK did the TEM imaging, EDS measurements, and corresponding analysis.

## Conflicts of interest

There are no conflicts to declare.

## Data availability

Data will be made available upon suitable request to the corresponding author.

## Acknowledgments

This work was partly carried out at the Micro and Nano Characterization Facility (MNCF), and National Nanofabrication Center (NNfC) located at CeNSE, IISc Bengaluru, funded by NPMAS-DRDO and MCIT, MeitY, Government of India; and benefitted from all the help and support from the staff. P.N. acknowledges the Start-up grant from IISc, Infosys Young Researcher award, and ANRF-SERB grant: CRG/2022/003506.